\documentclass[aps,prd,twocolumn]{revtex4-1}

\usepackage{graphicx}  
\graphicspath{ {images/} } 
\usepackage{latexsym}
\usepackage{slashed}
\usepackage{dcolumn}   
\usepackage{bm}        
\usepackage{amsmath}
\usepackage{amssymb}   
\usepackage{longtable}
\usepackage{float}
\newcommand{\be}{\begin{equation}}
\newcommand{\ee}{\end{equation}}
\newcommand{\bea}{\begin{eqnarray}}
\newcommand{\eea}{\end{eqnarray}}
\newcommand{\bma}{\begin{matrix}}
\newcommand{\ema}{\end{matrix}}

\newcommand{\bml}{\begin{mathletters}}
\newcommand{\eml}{\end{mathletters}}
\newcommand{\bes}{\begin{subequations}} 
\newcommand{\ees}{\end{subequations}}

\newcommand{\bi}{\begin{itemize}}
\newcommand{\ei}{\end{itemize}}

\begin{document}
\title{A non-vanishing neutrino mass and the strong CP problem: A new solution from the perspective of the EW-$\nu_R$ model}
\author{P. Q. Hung}
\affiliation{Department of Physics, University of Virginia,
Charlottesville, VA 22904-4714, USA}

\date{\today}

\begin{abstract}
The EW-$\nu_R$ model was constructed to provide a scenario in which right-handed neutrinos are non-sterile and have masses proportional to the electroweak scale, providing an opportunity to test the seesaw mechanism at the LHC. What was hidden in the model until recently is the chiral symmetry which helps solve the strong CP problem by using it to rotate $\theta_{QCD}$ to zero. It turns out the contribution from the electroweak sector to the effective $\bar{\theta}$ is proportional to the light neutrino masses and is naturally small, satisfying the constraint coming from the present absence of the neutron electric dipole moment, and without the need for an axion.\\

Talk presented at the APS Division of Particles and Fields Meeting (DPF 2017),
July 31-August 4, 2017, Fermilab. C170731
\end{abstract}

\pacs{}\maketitle

\section{Introduction}

The non-perturbative QCD vacuum is very complicated. 't Hooft \cite{thooft} has shown us that the proper gauge-invariant vacuum to use is the so-called $\theta$-vacuum:
\be
|\theta \rangle = \sum_{n} \exp (-\imath n \theta) | n \rangle \,. 
\ee
where $n$ is a "winding number".
This induces an Effective Lagrangian: $\mathcal{L}_{eff} = \mathcal{L}_{QCD+...} + \theta_{QCD} \, (g_{3}^2/32 \pi^2) G_{a}^{\mu \nu} \tilde{G_{\mu \nu}^{a}}$. The last term is CP violating! What would be the problem with that extra CP-violating term? It turns out that it contributes to the neutron electric dipole moment in the form which was worked out almost four decades ago by Crewther, Di Vecchia, Veneziano and Witten, \cite{crewther} namely
\be
d_n \approx 5.2 \times 10^{-16} \theta e-cm \,,
\ee
where more recent computations gave a lower value: 2.6 instead of 5.2. Experimentally \cite{ndipole}, the constraint is
\be
|d_n| < 2.9 \times 10^{-26} e-cm \,.
\ee
This leads to an upper bound on $\theta$:
\be
\theta < 10^{-10} \,.
\ee
Why $\theta$ should be so small is known as the strong CP problem. 

From Jackiw and Rebbi \cite{jackiw}, we learned that, under a $U(1)$ chiral transformation $\exp(\imath \alpha \tilde{Q}_5)$ ($\tilde{Q}_5$ is the axial charge), the $\theta$-vacuum changes
\be
\exp(\imath \alpha \tilde{Q}_5) |\theta \rangle = |\theta + \alpha \rangle \,.
\ee
If there is such a chiral symmetry, one can rotates $\theta =\theta_{QCD}$ to zero, namely $\exp(-\imath \theta_{QCD} \tilde{Q}_5) |\theta_{QCD} \rangle = |0 \rangle$ since all $\theta$-vacua are now equivalent. Thee is no more CP violation in the strong interactions! This is part of the Peccei-Quinn \cite{PQ} solution to the strong CP problem. 
The story is not so simple however. The diagonalization of the quark mass matrices introduces a chiral $U(1)$ transformation which further changes the $\theta$-vacuum by an amount $Arg Det M$ so that, in total, one has
\be
\label{thetabar}
\bar{\theta} = \theta_{QCD} + Arg Det M \,.
\ee
From Eq.~(\ref{thetabar}), even if one finds a chiral symmetry to rotate away $\theta_{QCD}$, it does not guarantee that $Arg Det M$ can be small enough so as to satisfy the experimental bound. Peccei and Quinn  \cite{PQ} proposed a famous solution by introducing a spontaneously broken global symmetry $U(1)_{PQ}$ which {\em dynamically} drives $\bar{\theta} $ to zero, upon the introduction of a dynamical field, the axion \cite{axion}. This is because, in general $Det M$ can be complex and, in addition, the VEV of the scalar field that breaks $U(1)_{PQ}$ can, in general, be large enough so as to violate the bound.

Experimental searches for the Axion (in particular Beam dump: $K^+ \rightarrow \pi^+ a$) turned out to be negative which led to the development of the concept of an Invisible axion \cite{invisible}. Severe constraints on the Axion from astrophysics although it is still considered to be a DM candidate. Because the present absence of an evidence for the axion, several axion-less scenarios were proposed \cite{axionless}.

It turns out that the EW-$\nu_R$ model \cite{nur} has all the necessary ingredients to provide a new solution to the strong CP problem.
\section{The EW-$\nu_R$ model \cite{nur}: A brief summary} 

\bi
\item It has the needed chiral symmetry to rotate away $\theta_{QCD}$.
\item The VEV of the scalar field which breaks that extra chiral symmetry is proportional to the neutrino mass and guarantees that $Arg Det M$ is sufficiently small so as to satisfy the bound on the neutron electric dipole moment. In fact, $Arg Det M \rightarrow 0$ as $m_{\nu} \rightarrow 0$. 
\item There is no need for an axion.
\ei
What is the EW-$\nu_R$ model and what does it do?

The EW-scale $\nu_R$ model is a model in which $\nu_R$'s are  NON-STERILE and have  EW-scale masses. As a result, they can be  detected at the LHC and the seesaw mechanism can be directly tested!
\bi
\item It does not require a new gauge group: It's the same old  $SU(3)_c \times SU(2)_W \times U(1)_Y$! (The subscript $W$ will be clarified below.)
\item It requires: ( One starts out with one family for simplification. One obtains similar results for 3 generations.)
\bi
\item Fermions: SM:  $(q_L, l_L,u_R, d_R, e_R)$; Mirror: $(q^{M}_R, l^{M}_R,u^{M}_L, d^{M}_L, e^{M}_L)$. Here the letters $q$ and $l$ stand for quark and lepton $SU(2)$ doublets.
\item Scalars: Doublets: $\Phi_2$, $\Phi_{2M}$; Triplets: $\tilde{\chi}$, $\xi$; Singlet: $\phi_S$.
\item The model contains an additional global symmetry: $U(1)_{SM} \times U(1)_{MF}$ \cite{nur,125}, under which the various fields transform as follows
\bea
U(1)_{SM}: &&\Phi_2 \rightarrow e^{\imath \alpha_{SM}} \Phi_2 \\ \nonumber
                 && (q_{L}^{SM}, l_{L}^{SM}) \rightarrow e^{\imath \alpha_{SM}} (q_{L}^{SM}, l_{L}^{SM}) 
\eea
 \bea
U(1)_{MF}: &&\Phi_{2M} \rightarrow e^{\imath \alpha_{MF}} \Phi_{2M} \\ \nonumber
                 && (q_{R}^{M}, l_{R}^{M}) \rightarrow e^{\imath \alpha_{MF}} (q_{R}^{M}, l_{R}^{M}) 
\eea

\bea
 \phi_S & \rightarrow& e^{-\imath ( \alpha_{MF} - \alpha_{SM})} \phi_S  \\ \nonumber
\tilde{\chi} & \rightarrow & e^{2 \imath \alpha_{MF}} \tilde{\chi}
\eea

All other fields are singlets under $U(1)_{SM} \times U(1)_{MF}$.

\item The additional global symmetry: $U(1)_{SM} \times U(1)_{MF}$ is needed to prevent, at tree-level, Yukawa couplings of the form: $\bar{f}_L \Phi_{2M} f_R$, $\bar{f}_{R}^{M}\Phi_{2} f_{L}^{M}$, $l_{L}^{T} \sigma_2 \tilde{\chi} l_{L}$. (This last one is forbidden if one does not want a large Majorana mass for left-handed neutrinos.)

\item Right-handed Majorana neutrino masses: $g_M \, \nu_{R}^{T}\,  \sigma_{2} \, \nu_R \chi^{0}$. With $\langle \chi^{0} \rangle = v_M$, one obtains $M_R = g_M\,v_M $ ($v_M \sim O(\Lambda_{EW})$).

\item Neutrino Dirac mass: ${\cal L}_S = g_{Sl} \, \bar{l}_{L}\, \phi_S \, l^{M}_{R} + H.c. $ with $\langle \phi_S \rangle = v_S$. One obtains  $m_D = g_{Sl} \, v_S $. 

\item Seesaw: Light neutrino (mostly left-handed) mass: $m_\nu \sim m_D^2/ M_R$; Heavy neutrino (mostly right-handed) mass: $M_R$. Numerically, one expects $m_D \sim O(100 \, keV)$ and $M_R \sim O(\Lambda_{EW})$.

\item A few remarks are in order here. As one can glimpse from the above discussion, the test of the seesaw mechanism can be completely performed at the LHC: $M_R$ through the discovery of EW-scale right-handed neutrinos and $m_D$  through the decays of mirror fermions into SM fermions and the Higgs singlet $\phi_S$ (missing energy). I will briefly come back to this at the end of the talk.

\ei

\ei

\section{The EW-$\nu_R$ model's perspective on the strong CP problem}

The Yukawa interactions of interest are
\bea
{\cal L}_{mass}&=& g_u \bar{q}_L \tilde{\Phi}_2 u_R + g_d \bar{q}_L \Phi_2 d_R \\ \nonumber
&& + g_{u^M} \bar{q^M}_R \tilde{\Phi}_{2M} u^M_L + g_{d^M} \bar{q^M}_R \Phi_{2M} d^M_L +H.c.
\eea
\bea
{\cal L}_{mixing} &=& g_{Sq}\bar{q}_L \phi_S \bar{q^M}_R + g_{Su}\bar{u}^M_L \phi_S  u_R  \\ \nonumber
&&+ g_{Sd}\bar{d}^M_L \phi_S  d_R + H.c.
\eea

$U(1)_{SM} \times U(1)_{MF}$ contains the chiral symmetries $U(1)_{A,SM} \times U(1)_{A,MF}$. The Yukawa interactions are invariant under the chiral transformations $q \rightarrow \exp(\imath \alpha_{SM} \gamma_5) q; q^M \rightarrow \exp(\imath \alpha_{MF} \gamma_5) q^M; \phi_S \rightarrow \exp(-\imath (\alpha_{SM}+ \alpha_{MF})) \phi_S $.
Under these chiral rotations: $\theta_{QCD} \rightarrow \theta_{QCD} -(\alpha_{SM} + \alpha_{MF})$. and it can be rotated to zero! This is like the first part of the Peccei-Quinn program. We are now left with the question of what to do with $Arg Det M$ (actually $\theta_{Weak} = Arg Det ({\cal M}_u {\cal M}_d) $). Is it sufficiently small? This has been answered in \cite{hungCP} and repeated here.

The mass matrices (for one family) are given by:
\bea
{\cal M}_u=\left( \begin{array}{cc}
	m_u & |g_{Sq} | v_S \exp(\imath \theta_q) \\
	 |g_{Su} | v_S  \exp(\imath \theta_u)& M_u
	\end{array} \right) \,,
\eea
	
\bea
{\cal M}_d=\left( \begin{array}{cc}
	m_d& |g_{Sq} | v_S \exp(\imath \theta_q) \\
	 |g_{Sd} | v_S  \exp(\imath \theta_d)& M_d
	\end{array} \right) \,.
\eea

Straightforward calculations give:
\be
\bar{\theta} =\theta_{Weak} \approx \frac{-(r_u \sin(\theta_q + \theta_u) 
+ r_d \sin(\theta_q + \theta_d))}{1- r_u \cos(\theta_q + \theta_u)-r_d \sin(\theta_q + \theta_d)} \,,
\ee
where
\be
r_u = \frac{|g_{Sq}||g_{Su}| v_S^2}{m_u M_u} = (\frac{|g_{Sq}||g_{Su}|}{g_{Sl}^2})(\frac{m_D^2}{m_u M_u}) \,,
\ee
and
\be
r_d= \frac{|g_{Sq}||g_{Sd}| v_S^2}{m_d M_d}=( \frac{|g_{Sq}||g_{Sd}|}{g_{Sl}^2})(\frac{m_D^2}{m_d M_d})
\ee

One can rewrite $\theta_{Weak}$ as
\be
\theta_{Weak} \approx -(r_u \sin(\theta_q + \theta_u) + r_d \sin(\theta_q + \theta_d)) \,.
\ee
With $r_{u,d} \propto v_S^2 \propto m_D^2$, $\theta_{Weak} \rightarrow 0$ as $v_S \rightarrow 0$ regardless of the CP-violating phases $\theta_q + \theta_{u,d}$. But $m_\nu \sim m_D^2/ M_R \neq 0$ and is tiny! One expects $\theta_{Weak}$ to be very small also! We do not need to drive $\theta_{Weak}$ dynamically to zero! This means that the strong CP problem, at least within the perspective of the EW-$\nu_R$ model, can be solved without the need for an axion.

Putting in some reasonable numbers, one gets
\be
\theta_{Weak} < -10^{-8}\{(\frac{|g_{Sq}||g_{Su}|}{g_{Sl}^2})\sin(\theta_q + \theta_u)  
+ (\frac{|g_{Sq}||g_{Sd}|}{g_{Sl}^2})\sin(\theta_q + \theta_d)\}
\ee
What are the implications of the above results and the EW-scale $\nu_R$ model itself?
\begin{itemize}
\item The EW-scale $\nu_R$ model:  (a) satisfies the EW-precision data, e.g. positive contributions to $S$ from mirror fermions get cancelled by negative contributions from triplet scalars \cite{vinh}; (b) Two very distinct scenarios Dr Jekyll and Mr Hyde that can accommodate, in terms of signal strengths, the 125-GeV scalar \cite{125}; (c)  Constraints from $\mu \rightarrow e \gamma$ \cite{muegamma} and $\mu2e$ conversion \cite{mu2e} imply $g_{Sl} < 10^{-4}$ which, in turns, implies that decays of mirror leptons at DISPLACED VERTICES!
\item To satisfy $\bar{\theta} < 10^{-10}$,  one requires $|g_{Sq}| \sim |g_{Su}| \sim |g_{Sd}| \sim 0.1 g_{Sl}$ which implies that decays of mirror quarks occur at DISPLACED VERTICES!
\item Another important collider implication: Like-sign dileptons from $\nu_R$ decays occur at DISPLACED VERTICES!.
\item Other implications are under investigation. Generalizations to the three-family case are being carried out. One expects the main conclusions to be similar to the one-family case
\end{itemize}

\section{Conclusions}
\begin{center}
What does the EW-scale $\nu_R$ model accomplish?
\end{center}
\begin{itemize}
\item Nielsen-Ninomiya theorem \cite{nielsen}: The SM cannot be put on the lattice. It would be tough to investigate the phase transition of the EW sector. The EW-scale $\nu_R$ model evades the N-N theorem and one can now study the phase transition on the lattice.
\item The EW-scale $\nu_R$ model provides a test of the seesaw mechanism at collider energies since  $\nu_R$'s are now NON-STERILE and "light"! Rich studies involving the search for the mirror sector at the LHC with in particular characteristic signals such as DISPLACED VERTICES underway.
\item There seems to be a collusion between neutrino physics and QCD to make Strong CP great again without the need for an axion!    
Stay tune!
\item Many of the topics discussed here are readily testable at the LHC.
\item Under investigation is the possibility that some scalar such as $\phi_S$ could act as a dark matter.

\end{itemize}

\end{document}